\documentclass[prl,showpacs,superscriptaddress,twocolumn,amsmath,amssymb]{revtex4}%
\usepackage{bm}
\newcommand{\iu}{{\rm i}}
\newcommand{\rmd}{{\rm d}}
\newcommand{\bfr}{{\mathbf r}}
\newcommand{\bfq}{{\mathbf q}}
\newcommand{\bfphic}{\bm{\psi}_{\rm c}}
\newcommand{\bfphizero}{\bm{\psi}_0}

\usepackage{graphicx}
\usepackage{epsfig}
\usepackage{amsmath}
\usepackage{graphics}

\begin{document}

\title{Dynamical Density Fluctuation of Superfluids near Critical Velocities}
 
\author{Yusuke Kato} 
\affiliation{Department of Basic Science, The University of Tokyo, Tokyo, 153-8902, Japan}
\author{Shohei Watabe}
\altaffiliation{Present Address: 
Department of Physics, Keio University, 3-14-1 Hiyoshi, Kohoku-ku, Yokohama 223-8522, Japan}
\altaffiliation{CREST(JST), 4-1-8 Honcho, Kawaguchi, Saitama, 332-0012, Japan}
\affiliation{Institute of Physics, Department of Physics, The University of Tokyo, Tokyo 153-8902, Japan}


\begin{abstract}
We propose a stability criterion of superfluids in condensed Bose-Einstein systems, which incorporates the spectral function or the autocorrelation function of the local density. Within the Gross-Pitaevskii-Bogoliubov theory, we demonstrate the validity of our criterion for the soliton-emission instability, with use of explicit forms of zero modes of the Bogoliubov equation and a dynamical scaling near the saddle-node bifurcation. We also show that the criterion is applicable to the Landau phonon instability and the Landau roton instability within the single-mode approximation.  
\end{abstract}


\maketitle
%

{\it Introduction}. 
Studies of the critical velocity,  above which a superfluid becomes unstable, 
can trace its history back to the discovery of superfluidity in $^4$He \cite{wilks67}. 
The critical velocity is determined by either of the following, whichever yields the smaller critical velocity: 
the instability of the excitation spectrum with respect to a flowing superfluid (the so-called Landau instability \cite{landau41}) 
or the emission of topological defects such as quantized vortices \cite{feynman55,anderson66}. 

The realization of Bose-Einstein condensation in cold atoms has renewed interest in this issue. 
The critical velocities in cold atoms have been measured by moving a blue-detuned laser beam in a Bose-Einstein condensate (BEC) \cite{raman99,onofrio00,inouye01}, where superfluidity was found to break down with the emission of vortices at the velocities much smaller than the Landau critical values. 
Those experimental results \cite{raman99,onofrio00} are consistent with numerical results of the time-dependent Gross-Pitaevskii \cite{gross61,pitaevskii61} (GP) equations \cite{frisch92,winiecki99}.

From the viewpoint of non-linear physics, the instability of superfluidity found in \cite{frisch92,winiecki99} was identified as a saddle-node bifurcation \cite{pomeau93,huepe00}, where a stable and an unstable steady solutions for the GP equation merge \cite{guckenheimer83}. 
As a simpler analog of the vortex-emission instability, the instability of the one-dimensional superflow against a penetrable potential barrier has been intensively studied \cite{hakim97,baratoff70, pavloff02} within the GP equation; superfluidity breaks down with the emission of solitons much below the Landau critical velocity at a saddle-node bifurcation \cite{hakim97,pham02}. 

If we could find a common aspect between the two different kinds of instability: the Landau instability and the saddle-node bifurcation, it would provide a crucial step toward a coherent understanding of various kinds of instability in superfluids. 
In this Letter, we focus on dynamical density fluctuations (DDFs) and show that the DDFs of superfluids are enhanced near the critical velocity both in the Landau instability and soliton emission instability. 
On the basis of this finding, we propose a criterion for stability of superfluids\cite{priorpublication}.

{\it Model}. 
We quantify DDFs through the spectral function of the local density, 
since the dynamical structure factor, usually denoted as $S({\mathbf q},\omega)$ in the literature, 
is not appropriate to measure density fluctuations in the presence of obstacles. 
The spectral function of the local density is given by 
\begin{equation}
I({\mathbf r},\omega)=\sum_l |\langle l |\hat{\psi}^\dagger({\mathbf r})\hat{\psi}({\mathbf r})|{\rm g}\rangle|^2\delta(\omega -E_l+E_{\rm g}).
\label{eq: ldsf}
\end{equation}
Here, $|{\rm g}\rangle$ is the state vector of the ground state or a stable superflow state, the energy of which is denoted by $E_{\rm g}$. 
$E_l$ denotes the energy of an excited state $l$. 
The summation runs over the excited states.

We consider Bose systems with a short-range repulsive two-body interaction at zero temperature. The $x$-axis is taken to be the direction of the supercurrent. 
We introduce a penetrable and short-ranged potential barrier $U_{\rm ex}(x)$, which is localized along the $x$-direction around $x=0$ (independent of $y$ and $z$). 
The shape of the barrier is similar to that shown in Fig.~1 a) of \cite{engels07}. 
We treat the condensate and excitations within the scheme of the GP \cite{gross61,pitaevskii61} and the Bogoliubov \cite{bogoliubov47} equations, respectively. The field operator is thus given by
$
 \hat{\psi}({\mathbf r},t)=\Psi({\mathbf r})+\sum_j[u_j({\mathbf r})\hat{a}_j {\rm e}^{-\iu\varepsilon_j t}
  -v^{*}_j({\mathbf r})\hat{a}^{\dagger}_j {\rm e}^{\iu\varepsilon_j t}]. 
$
The function $\Psi({\mathbf r})$ represents the wavefunction of the condensate and satisfies the stationary GP equation: 
\begin{equation}
\left[-\frac{\textrm{\boldmath $\nabla$}^2}{2}+U_{\rm ext}(x)-\mu+|\Psi({\mathbf r})|^2\right]\Psi({\mathbf r})=0 \label{Gross-Pitaevskii equation}.
\end{equation}
The symbols $\hat{a}_j$ and $\hat{a}_j^\dagger$ denote, respectively, the annihilation and creation operators of a one-particle state $j$. The wavefunctions $u_j({\mathbf r})$ and $v_j({\mathbf r})$ of the state $j$ satisfy the Bogoliubov equation:
  \begin{equation}
{\cal L}  \left(
  \begin{array}{c}
  u_j({\mathbf r}) \\
  v_j({\mathbf r}) \\
  \end{array} 
  \right)
  =\varepsilon_j
  \left(
  \begin{array}{c}
  u_j({\mathbf r}) \\
  v_j({\mathbf r}) \\
  \end{array} 
  \right),\quad
{\cal L}\equiv\left(
  \begin{array}{cc}
  \hat{K} & -\Psi^2 \\
  (\Psi^{*})^2 &-\hat{K} \\
  \end{array} 
  \right), 
\label{eq: Bogoliubov}
  \end{equation}  
\begin{equation}
\hat{K}\equiv -\frac{\textrm{\boldmath $\nabla$}^2}{2}+U_{\rm ext}(x)-\mu+2|\Psi({\mathbf r})|^2, 
\label{eq: K}
\end{equation}
and also satisfy the ortho-normal condition
$\int\rmd \bfr[u_j({\mathbf r})u^{*}_k({\mathbf r})
  -v_j({\mathbf r})v^*_k({\mathbf r})]=\delta_{jk}$, 
and 
$\int\rmd \bfr[u_j({\mathbf r})v_k({\mathbf r})
  -v_j({\mathbf r})u_k({\mathbf r})]
  =0.
$ 
We have used dimensionless units in (\ref{Gross-Pitaevskii equation}) and (\ref{eq: Bogoliubov}). 

In the presence of a potential barrier $U_{\rm ex}(x)$, 
we consider (\ref{Gross-Pitaevskii equation}) under the boundary condition
\begin{equation}
\Psi(x\rightarrow\pm\infty) = \exp\Big[\mathrm{i}\Big(V x+\varphi_{\pm}\Big)\Big]. \label{eqGPasymp}
\end{equation}
The quantity $\varphi\equiv \varphi_+ -\varphi_-$ represents the phase difference between the two condensates separated by the potential barrier. $V$ represents the superfluid velocity far away from the potential barrier. The relation between $V$ and $\varphi$ corresponds to the Josephson relation \cite{baratoff70,danshita06} and its typical profile 
is given, e.g. in Fig.~2 of \cite{baratoff70}. 
For chemical potential $\mu$, the relation $\mu=1+V^2/2$ follows from the boundary condition (\ref{eqGPasymp}). 

When $U_{\rm ex}(x)=U_0 \delta(x)$ with $U_0>0$, the following facts are known from the analysis of Refs.~\cite{hakim97,pham02,pavloff02}:
\begin{enumerate}
\item
The critical velocity $V_{\rm c}$ is smaller than the Landau critical velocity(=1).  
\item
When $0 \le V<V_{\rm c}$, a stable and an unstable steady solutions to (\ref{Gross-Pitaevskii equation}) exist. 
\item
When $V\rightarrow V_{\rm c}-0$, the stable and the unstable steady solutions to (\ref{Gross-Pitaevskii equation}) merge.
\end{enumerate} 
As for $V>V_{\rm c}$, no steady solutions to (\ref{Gross-Pitaevskii equation}) with dissipationless flows exist. When $V$ is slightly above $V_{\rm c}$, the emission of gray solitons is observed in numerical calculations of the time-dependent Gross-Pitaevskii equation \cite{hakim97}. 
When $U_{\rm ex}(x)$ is a rectangular barrier, 
a Josephson-type $V$-$\varphi$ relation was also derived in \cite{danshita06}. A rectangular barrier describes qualitatively generic properties 
of a penetrable short-range barrier. 
We thus assume the $V$-$\varphi$ relation for a generic short-range barrier to be a Josephson-type relation, from which (a), (b) and (c) follow. 

{\it DDF in the presence of a potential barrier}. 
The spectral function $I(x,\omega)$ at low $\omega$ is derived via the method used in \cite{takahashi09}; first we find the zero modes of the Bogoliubov equation (\ref{eq: Bogoliubov}) and next obtain the solutions of (\ref{eq: Bogoliubov}) 
in the form of the series with respect to energy. 
On the basis of the low energy asymptotics of the matrix element and the density of excited states, 
we discuss the details of $I(x,\omega)$ at low $\omega$. 

In the Bogoliubov theory, the spectral function (\ref{eq: ldsf}) reduces to 
$
I({\mathbf r},\omega)=\sum_{j}|\Psi^*(\bfr)u_j(\bfr)-\Psi(\bfr)v_j(\bfr)|^2\delta(\omega-\varepsilon_j). 
$
For simplicity, we first discuss the case $d=1$. 
When the volume $\Omega$ of the system is sufficiently large, a typical value $\Delta\varepsilon$ of the level-spacing $|\varepsilon_{j+1}-\varepsilon_j|$ is of the order of $\Omega^{-1}$ for small $\varepsilon_{j}$.

We here introduce a function $M(x,\omega)$ satisfying 
\begin{eqnarray}
&&M(x,\omega)=\Omega |\Psi^*(x)u_j(x)-\Psi(x)v_j(x)|^2,\label{eq: cond-M-2}
\end{eqnarray}
when $\omega=\varepsilon_j$ for each $j$. 
Further we require that $M(x,\omega)$ varies over the energy scale much larger than $\Delta \varepsilon$, i.e.
\begin{eqnarray}
&&\Big|\partial M(x,\omega)/\partial \omega\Big|\Delta\varepsilon\ll
\Big|M(x,\omega)\Big|.\label{eq: cond-M-1} 
\end{eqnarray}
In terms of $M(x,\omega)$, the spectral function is expressed as
\begin{equation}
I(x,\omega)=M(x,\omega)\Omega^{-1}\sum_j\delta(\omega-\varepsilon_j)\equiv M(x,\omega)D(\omega).
\label{eq: MD}
\end{equation}
Here $D(\omega)$ denotes the one-particle density of states for the Bogoliubov mode, which becomes a constant at low $\omega$ for $d = 1$ in the thermodynamic limit ($\Omega\rightarrow \infty$). 

In order to discuss $M(x,\omega)$ at low $\omega$, 
we first recall that 
$
\left(u(x),v(x)\right)^{\rm t}=
\left(\Psi(x),\Psi^*(x)\right)^{\rm t}\equiv\bm{\psi}_0
$
is a solution of (\ref{eq: Bogoliubov}) with zero eigenvalue (zero mode) for $V\le V_{\rm c}$\cite{fetter72}.
Next, one can verify from the following observation that 
\begin{equation}
\left(u(x),v(x)\right)^{\rm t}=
\left(\partial \Psi(x)/\partial \varphi,-\partial \Psi^*(x)/\partial \varphi\right)^{\rm t}\equiv \bfphic
\label{eq: zeromode2}
\end{equation}
is an additional zero mode of (\ref{eq: Bogoliubov}) at $V=V_{\rm c}$\cite{pham02,takahashi09}.
Regarding $V$ as a function of $\varphi$ and taking the derivative of (\ref{Gross-Pitaevskii equation}) with respect to $\varphi$, we obtain 
\begin{equation}
K(\partial \Psi/\partial \varphi)+\Psi^2(\partial\Psi^*/\partial\varphi)=V(\rmd V/\rmd \varphi)\Psi.
\end{equation}
This equation and its complex conjugation are put together as
  \begin{equation}
{\cal L}\bfphic  
  =V(\rmd V/\rmd\varphi)\bfphizero.
\label{eq: zeromode2equation}
\end{equation}
When $V=V_{\rm c}$, $V$ is maximal and hence $\rmd V/\rmd \varphi=0$. 
The right-hand side in (\ref{eq: zeromode2equation}) then vanishes and the proof is completed. 
 
When $V<V_{\rm c}$, the wave function for a state $j$ with a small $\varepsilon_j$ can be shown to have the form\cite{takahashi09,watabethesis}: 
\begin{equation}
\left(
u_j(x),
v_j(x)\right)^{\rm t}
=(\varepsilon_j\Omega)^{-\frac12}\left\{
c\bm{\psi}_0(x)
+c\varepsilon_j \bm{\psi}_1(x)+{\cal O}(\varepsilon_j^2)\right\}.
\label{eq: zeroenergysolvlvc}
\end{equation}
Here $\bm{\psi}_1(x)$ is the solution of ${\cal L}\bm{\psi}_1=\bm{\psi}_0$ which does not diverge exponentially at large $|x|$.
When $V=V_{\rm c}$, on the other hand, we have \cite{pham02,takahashi09,watabethesis}
\begin{equation}
\left(
u_j(x),v_j(x)\right)^{\rm t}
=
(\varepsilon_j\Omega)^{-\frac12}\left\{ c
\bm{\psi}_0 (x)+c'\bm{\psi}_{\rm c}(x)
+{\cal O}(\varepsilon_j)\right\}.
\label{eq: uvlimitVc}
\end{equation}
Here $c$ and $c'$ are constants.

$\bfphizero$ represents a phase fluctuation \cite{takahashi09} and hence does not contribute to density fluctuations, while $\bfphic$ does\cite{takahashi09}. 
Correspondingly, the matrix element $M(x,\omega)$ behaves as
\begin{equation}
M(x,\omega)
\rightarrow \bar{M}(x,\omega)\equiv
\left\{
\begin{array}{ll}
\omega f(x) &\quad V<V_{\rm c}, \\
 & \\
\frac{c'^2}{\omega}\left|\frac{\partial |\Psi(x,\varphi)|^2}{\partial \varphi}\right|^2 &\quad V=V_{\rm c}, 
\end{array}
\right. 
\label{eq: matrixelementbehavior}
\end{equation}
at low $\omega$. Here $f(x)$ represents the contribution from ${\mathbf \psi}_1$. The low $\omega$ behavior of $I(x,\omega)$ for $d=1$ can be deduced as $\bar{M}(x,\omega)D(\omega)$ from (\ref{eq: MD}) and (\ref{eq: matrixelementbehavior}).

The generalization to $d=2$ and 3 is straightforward as outlined below~\cite{katowatabeunpub}; we classify the set of the eigenstate $j$ with the angle $\phi$ 
between the wave vector and the potential wall, and introduce $M(x,\omega,\phi)$ satisfying (\ref{eq: cond-M-2}) and (\ref{eq: cond-M-1}) for $j$ with an angle $\phi$. At low $\omega$, we can show that $M(x,\omega,\phi)$ becomes independent of $\phi$ and that it approaches $\bar{M}(x,\omega)$. Consequently, $I(x,\omega)\rightarrow \bar{M}(x,\omega)D(\omega)$ at low $\omega$.  Since $D(\omega)\propto \omega^{d-1}$ at low $\omega$ for the Bogoliubov mode in $d(=1,2,3)$-dimension, we obtain 
\begin{equation}
I(x,\omega)
\rightarrow 
\left\{
\begin{array}{ll}
\omega^{d} f_d(x,V) &\quad V<V_{\rm c}, \\
 & \\
\nu_d \omega^{d-2}\left|\frac{\partial |\Psi(x,\varphi)|^2}{\partial \varphi}\right|^2 &\quad V=V_{\rm c}, 
\end{array}
\right. 
\label{eq: Ixomegabehavior}
\end{equation}
at low $\omega$. Here $f_d(x,V)$ is a function of $x$ and $\nu_d$ is a constant. 

In (\ref{eq: Ixomegabehavior}), we note that the exponent in $\omega$ jumps from $d$ to $d-2$ when $V$ approaches $V_{\rm c}$ from below. This behavior can be explained by 
a dynamical scaling relation near the saddle-node bifurcation \cite{pham02,huepe00}. Near the saddle-node bifurcation, there is a characteristic frequency $\omega^*$, which scales as $\omega^*\propto |V-V_{\rm c}|^\frac12$. 
In \cite{pham02}, it was shown that the decay rate $\Gamma$ of an unstable steady state slightly below $V_{\rm c}$ and the emission rate $\Gamma'$ of gray soliton slightly above $V_{\rm c}$ follow the scaling relations equivalent to $\Gamma,\Gamma'\propto |V-V_{\rm c}|^\frac12$ \cite{noteSN}. 
We thus anticipate that $\omega^*$ yields a crossover behavior in the $\omega$-dependence of $I(x,\omega)$ near $V_{\rm c}$. 

\begin{figure}
\begin{center}
\includegraphics[%
  width=0.53\linewidth,
  ]{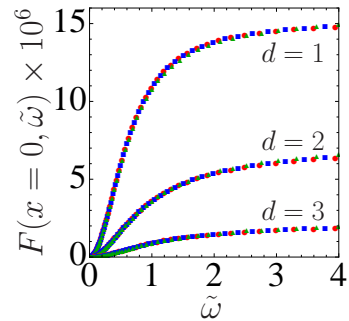}
\end{center}
\caption{(Color online)
Scaling relation near the saddle-node bifurcation for dimensions $d=1, 2$ and $3$. 
The scaled spectral functions $F(x=0,\tilde{\omega}=\omega|V-V_{\rm c}|^{-1/2}) = \omega^{2-d}I(x=0,\omega)$ near the critical current 
are shown as a function of scaled frequency $\tilde{\omega}$. 
The critical velocity $V_{\rm c}$ in dimensionless form is given by $V_{\rm c} = 0.049753 \cdots$ for $U_{\rm ex}(x)=U_0\delta(x)$ with $U_0=10$. 
Circle (red), square (blue), and triangle (green) points are data for $V = 0.04975, 0.0497$ and $0.049$, respectively. 
}
\label{fig: snscaling}
\end{figure}

Accordingly, we put $I(x,\omega)$ in the scaling form:
\begin{equation}
I(x,\omega)=\omega^{d-2}F(x,\omega|V-V_{\rm c}|^{-\frac12}),\label{eq: snscaling}
\end{equation}
 where the asymptotic behaviors of $F(x,\tilde{\omega}\equiv \omega|V-V_c|^{-\frac12})$ are determined from 
(\ref{eq: Ixomegabehavior}) as
\begin{equation}
F(x,\tilde{\omega})\rightarrow \left\{
\begin{array}{cc}
\tilde{\omega}^2 \tilde{f}_d(x)&\mbox{ for }\tilde{\omega}\rightarrow 0, \\
\nu_d\Big|\frac{\partial |\Psi(x,\varphi)|^2}{\partial \varphi}\Big|^2&\mbox{ for }\tilde{\omega}\rightarrow \infty, \\
\end{array}\right. 
\label{eq: Fxomega}
\end{equation}
with $\tilde{f}_d(x)=(V_{\rm c}-V) f_d(x,V)$. 
The scaling relation (\ref{eq: snscaling}) with (\ref{eq: Fxomega}) shows that 
$I(x,\omega)$ is enhanced near $V=V_{\rm c}$ at low $\omega$.
We have confirmed this scaling relation numerically. 
Figure~\ref{fig: snscaling} shows the scaling plot $\omega^{2-d} I(x=0,\omega)$ vs. $\tilde{\omega}$ for $U_{\rm ex}(x)=U_0\delta(x)$ with $U_0=10$ near $V_{\rm c}$. We see that for each dimension, the data for various values of $V$ near $V_{\rm c}$ collapse onto a single curve. 

In deriving the scaling relation (\ref{eq: snscaling}) with (\ref{eq: Fxomega}), the following facts were crucial: 
(I) the zero mode which couples to density fluctuations exits only at $V=V_{\rm c}$, 
and 
(II) the characteristic frequency $\omega^*$ scales as $|V_{\rm c}-V|^{1/2}$ near the saddle-node bifurcation. 
These two facts hold in superfluid Bose systems where the critical velocity is determined by a saddle-node bifurcation. 
We thus expect that a scaling relation similar to (\ref{eq: snscaling}) holds also in the case of a superflow around a disk \cite{frisch92,pomeau93,huepe00}, where the vortex-emission instability of superfluids has been identified with a saddle-node bifurcation. 

{\it DDF near the Landau critical velocity.}
We next show that the enhancement of the DDF near $V=V_{\rm c}$ also occurs in the Landau instability. 
To that end, we consider the DDF near the Landau critical velocity in spatially uniform systems within the single mode approximation (SMA) \cite{feynman56}. We assume that the excitation energy $\varepsilon_q$ in the absence of the superflow ($V=0$) depends only on $|{\mathbf q}|=q$ and has the form $\varepsilon_q\sim c_1 q+c_3 q^3+{\cal O}(q^5)$ with $c_1>0$ near $q=0$.  In the SMA, $\varepsilon_q$ is related to the dynamical structure factor $S(q,\omega)$ for $V=0$ via \cite{feynman56}
\begin{equation}
\varepsilon_q\sim \frac{q^2}{2S(q)},\quad S(q)=\int \rmd \omega S(q,\omega),
\end{equation}
from which $S(q,\omega)\sim (q^2/(2\varepsilon_q))\delta(\omega -\varepsilon_q)$ follows.  When $V\ne 0$, $S(\bfq,\omega)$ is given by
\begin{equation}
S({\mathbf q},\omega)\sim (q^2/(2\varepsilon_q))\delta(\omega -\varepsilon_q -V q_x). \label{eq: Sqomega}
\end{equation}
In spatially uniform systems, the spectral function $I(\bfr,\omega)=I(\omega)$ is related to the dynamical structure factor via $I(\omega)=(2\pi)^{-d}\int\rmd\bfq S(\bfq,\omega)$. From this and (\ref{eq: Sqomega}), we derive $I(\omega)$ at low $\omega$ in the following two cases.\\
(i) Landau phonon instability. When the Landau instability occurs near $q=0$ for $c_3>0$, 
the Landau critical velocity $V_{\rm c}$ is given by the sound velocity $c_1$. 
$I(\omega)$ behaves as
\begin{equation}
I(\omega)\propto \left\{
\begin{array}{ll}
\omega^{d} &\mbox{ for }\omega\ll \omega^* , 
\\
\omega^{(2d-3)/3} &\mbox{ for }\omega^*\ll \omega\ll \omega_{\rm l} , 
\end{array}
\right.
\label{eq: crossoverlandau}
\end{equation} 
with $\omega^*\sim (c_1-V)^{3/2}(c_3)^{-1/2}$ and $\omega_{\rm l}\sim (c_1-V)c_1^{1/2}c_3^{-1/2}$. \\
(ii) Landau roton instability. 
We consider a system where the single mode $\varepsilon_q$ has a phonon-roton spectrum in the absence of $V$. The critical velocity $V_{\rm c}$ is determined by demanding that the condition $\varepsilon_q -V_{\rm c} q=0$ be met for some wave vector ${\mathbf q}={\mathbf q}_{\rm c}$ which lies antiparallel to the superfluid velocity. 
Around ${\mathbf q}\sim 
{\mathbf q}_{\rm c}$, $\varepsilon_q -V_{\rm c} q$ can be expanded as $\sim q_{\rm c}(V_{\rm c}-V)+\Delta ({\mathbf q}-{\mathbf q}_{\rm c})^2$ 
with a positive coefficient $\Delta$. Near $V=V_{\rm c}$ and low $\omega$, $S({\mathbf q},\omega)$ is given by
\begin{eqnarray}
S({\mathbf q},\omega)&\sim& (q_{\rm c}^2/(2\varepsilon_{q_{\rm c}}))\delta(\omega -\Delta ({\mathbf q}-{\mathbf q}_{\rm c})^2-q_{\rm c}(V_{\rm c}-V))\nonumber\\
&&+(q/(2 c_1))\delta(\omega-c_1 q+V q_x).
\end{eqnarray}
In the right-hand side, the first and second terms represent roton and phonon contributions, respectively. 
As a result, $I(\omega)$ at low $\omega$ is given by 
\begin{equation}
I(\omega)\sim A \omega^{(d-2)/2}\theta(\omega-q_{\rm c}(V_{\rm c}-V))+B \omega^{d},\label{eq: LandaurotonIomega}
\end{equation}
where $A$ and $B$ are constants. The symbol $\theta$ denotes the Heaviside step function. 

Both (\ref{eq: crossoverlandau}) and (\ref{eq: LandaurotonIomega}) show that the DDFs are enhanced at low $\omega$ when the superfluid velocity approaches the Landau critical velocity. 
\\
{\it Stability criterion of superfluids.} 
The enhancement of the DDF near the critical velocity is a common aspect 
between a saddle-node bifurcation and the Landau phonon/roton instability. 
For those kinds of instability, 
the DDF of superfluids in $d$-dimensional BECs is explicitly summarized in the following unified way: 
\begin{equation}
\begin{array}{l}
I({\mathbf r},\omega)=\omega^\beta f(\bfr)+I'(\bfr,\omega), \\
\lim\limits_{\omega\rightarrow +0} \omega^{-\beta}I'(\bfr,\omega)=0 
\quad \mbox{ for }^\forall\bfr, 
\end{array}
\label{eq: criterion}
\end{equation} 
where the exponent $\beta$ satisfies 
\begin{equation}\beta=
\left\{
\begin{array}{lc}
\beta_{\rm c}(<d)& \quad V=V_{\rm c}, \\
d &\quad V<V_{\rm c}. \\
\end{array}
\right.\label{eq: beta}
\end{equation}
The function $f(\bfr)$ and the exponent $\beta_{\rm c}$ at $V=V_{\rm c}$ depend on the mechanism for destabilizing a superfluid state. 

The relation (\ref{eq: criterion}) can be also expressed in terms of the autocorrelation function of the local density. We introduce ^^ ^^ the coarse-grained local density" $\tilde{n}({\bfr}, t)=\int w_{a}(|\bfr -{\mathbf r'}|)\hat{\psi}^\dagger({\bfr}', t)\hat{\psi}({\bfr}', t)\rmd \bfr', $ in order to avoid the unessential divergence of the autocorrelation at the short-time limit.
The weight function $w_a(r)$ (with a coarse-graining radius $a$) satisfies
$
w_a(r)\sim 0$ for $r\gg a (>0)$, and $\int w_a(|\bfr|)\rmd\bfr=1. 
$
The long-time behavior of the autocorrelation function 
$
C(\bfr,t)=\langle \tilde{n}(\bfr,t)\tilde{n}(\bfr,0)+\tilde{n}(\bfr,0)\tilde{n}(\bfr,t)\rangle/2-\langle \tilde{n}(\bfr,0)\rangle^2
$
is determined by
the low $\omega$ property of $I(\bfr,\omega)$ and 
it follows that for large $t$,  
\begin{equation}
\begin{array}{l}
C({\mathbf r},t)=t^{-(\beta+1)}\tilde{f}(\bfr)+C'(\bfr,t), \\
\lim_{t\rightarrow \infty}C'(\bfr,t)t^{\beta+1}=0
\quad \mbox{ for }^\forall\bfr, 
\end{array}
\label{eq: criterion-Ct}
\end{equation}
with (\ref{eq: beta}). $\tilde{f}(\bfr)$ is a function of $\bfr$.
The relation (\ref{eq: criterion-Ct}) reveals another aspect of instability of a superfluid state; 
a qualitative change in density correlation in the long-time domain 
is a sign that the system has reached the critical velocity. 

We propose (\ref{eq: criterion}) and (\ref{eq: criterion-Ct}) as new criteria of stability of superfluids of BECs. 
The autocorrelation of local density would be more accessible than the spectral function in experiments of cold atoms. We thus expect that (\ref{eq: criterion-Ct}) is more practically useful than (\ref{eq: criterion}).   

{\it Conclusion}. 
We examined the dynamical density fluctuations on superfluids 
both in the presence of the potential barrier and spatially uniform systems.  
On the basis of results for the saddle-node bifurcation and for the Landau instability, we propose a dynamical criterion for stability of superfluids.

{\it Acknowledgments.} 
We thank D.~Takahashi, Y.~Nagai, M.~Kunimi, T.~Minoguchi, S.~Sasa, M.~Kobayashi and H.~Ohta for useful discussions. SW is grateful to G. Baym for his comments on the Landau instability. We thank A.~Tanaka for his critical reading of the manuscript. 
This work is supported by KAKENHI (21540352) from JSPS and (20029007) from MEXT in Japan. SW acknowledges support from Grant-in-Aid for JSPS Fellows (217751).


\begin{thebibliography}{99}
\bibitem{noteSN}Precisely speaking, the control parameter is not $V$ but the strength of potential barrier in \cite{pham02}.
\end{thebibliography}
\end{document}